\documentclass[final,3p,12pt]{elsarticle}
\usepackage{amsfonts}
\usepackage{amssymb}
\usepackage{amsmath}
\usepackage{graphicx}
\usepackage{epstopdf}

\journal{Physics Letters A}

\begin{document}

\begin{frontmatter}

\title{A Korteweg-de Vries description of dark solitons in polariton superfluids}

\author{R.~Carretero-Gonz{\'a}lez}
\address{Nonlinear Dynamical Systems Group,
Computational Sciences Research Center, and
Department of Mathematics and Statistics,
San Diego State University, San Diego, California 92182-7720, USA}

\author{J. Cuevas-Maraver}
\address{Grupo de F\'{i}sica No Lineal, Departamento de F\'{i}sica Aplicada I, 
Universidad de Sevilla. Escuela Polit\'{e}cnica Superior, C/ Virgen de \'{A}frica, 7, 41011-Sevilla, Spain
\\
Instituto de Matem\'{a}ticas de la Universidad de Sevilla (IMUS). Edificio Celestino Mutis. Avda. Reina Mercedes s/n, 41012-Sevilla, Spain}

\author{D.~J. Frantzeskakis}
\address{Department of Physics, National and Kapodistrian University of Athens, Panepistimiopolis, Zografos, Athens 15784, Greece}

\author{T.~P. Horikis}
\address{Department of Mathematics, University of Ioannina,
Ioannina 45110, Greece}

\author{P.~G. Kevrekidis}
\address{Department of Mathematics and Statistics, University of Massachusetts,
Amherst, Massachusetts 01003-4515 USA}

\author{A.~S.\ Rodrigues}
\address{Departamento de F\'{\i}sica/CFP, Faculdade de Ci\^{e}ncias, 
Universidade do Porto, R. Campo Alegre,
687 - 4169-007 Porto, Portugal}

\begin{abstract}

We study the dynamics of dark solitons in an incoherently pumped exciton-polariton condensate 
by means of a system composed by a generalized open-dissipative Gross-Pitaevskii equation 
for the polaritons' wavefunction and a rate equation for the exciton reservoir density. 
Considering a perturbative regime of sufficiently small reservoir excitations, we use 
the reductive perturbation method, to reduce the system to a Korteweg-de Vries (KdV) equation 
with linear loss. This model is used to describe the analytical form and the dynamics 
of dark solitons. We show that the polariton field supports decaying dark soliton
solutions with a decay rate determined analytically in the weak pumping regime. 
We also find that the dark soliton evolution is accompanied by a shelf, 
whose dynamics follows qualitatively the effective KdV picture. 

\end{abstract}


\end{frontmatter}

\section{Introduction}

The experimental realization of Bose-Einstein condensates (BECs) of exciton-polaritons 
\cite{exp1,exp2,exp3,exp4}, namely hybrid light-matter quasiparticles emerging in the 
regime of strong coupling, has triggered the emergence of an exciting research field, 
where both quantum and nonequilibrium dynamics are present~\cite{rmp1st,cc}. 
From a mathematical point of view, since these systems are intrinsically lossy and, hence, 
need to be continuously replenished, they are described by specifically tailored 
damped-driven variants of the Gross-Pitaevskii (GP) equation~\cite{wc,wc2,wc3,berl1,berl2} 
(see also discussion in the review~\cite{cc}). 
It is important to point out that the Hamiltonian variant of this model is widely used 
in the context of atomic BECs~\cite{peth,pita,siambook}, where it can successfully 
describe, under experimentally relevant conditions, the statics and dynamics of BECs, 
as well as a plethora of nonlinear phenomena emerging in this context~\cite{emergent} 
(see also the reviews~\cite{nonlin,djf,rom}). 
Similarly, in the context of polariton condensates, 
versions of an open-dissipative GP model describing an incoherently pumped polariton BEC 
coupled to the exciton reservoir~\cite{wc,wc2,wc3} were successfully used in the theoretical
description of a number of seminal experiments reporting, e.g., the formation of 
quantized vortices~\cite{vort1,vort2,vort3} (see also the
very recent work of Ref.~\cite{vort3a}) and dark solitons~\cite{ds1,ds2,ds3,ds4,ds5,ds6} 
(see also the experiment reported in Ref.~\cite{ds0} and the theoretical work 
of Ref.~\cite{wc3} related to experiments~\cite{ds1,ds2,ds3,ds4}).

Dark solitons and their dynamics in polariton superfluids, which is the theme of this paper, 
have been studied in various works~\cite{dsth1,dsth1b,dsth1c,dsthwe1,dsthwe2,ofy,dsthv1,dsthv2}. 
In particular, in Refs.~\cite{dsth1,dsth1b,dsth1c}, dark solitons were analyzed in polariton 
condensates coherently and resonantly driven by a pumping laser. On the other hand, 
in Refs.~\cite{dsthwe1,dsthwe2}, a simplified Ginzburg-Landau type model~\cite{berl1,berl2} 
was used to analyze one-dimensional and ring dark solitons, respectively, in the presence of 
nonresonant pumping. In the same case (of nonresonant pumping), 
and using the model of Refs.~\cite{wc,wc2,wc3}, which involves the coupling of polaritons 
to the exciton reservoir, dark polariton solitons were analyzed using an 
adiabatic approximation~\cite{ofy} and variational techniques~\cite{dsthv1,dsthv2}. 
Here, we consider the problem of dark polariton soliton dynamics, and adopt 
the model of Refs.~\cite{wc,wc2,wc3}, which is perhaps the most customary 
approach to describe an incoherently pumped exciton-polariton BEC. This model 
is a generalization of the open-dissipative GP equation for the macroscopic 
wavefunction of the polariton condensate, by couplind it to a rate equation 
for the exciton reservoir density. 
Assuming that this system is quasi one-dimensional (1D), e.g., a 1D microcavity~\cite{ds5} 
or a nanowire~\cite{nw}, and in the uniform pumping regime, we consider a 
perturbative approach for the reservoir excitations. 
Under these assumptions, we show that it is possible to study this system analytically, 
by employing the reductive perturbation method~\cite{rpm}. In particular, using this approach, 
we derive an effective Korteweg-de Vries (KdV) equation with linear loss;  
generalizing the pertinent result relevant to the Hamiltonian case of quasi-1D atomic BECs, 
where small-amplitude dark solitons obey an effective KdV equation (see the 
reviews~\cite{nonlin,djf} and references therein). In the small-amplitude limit 
under consideration, we show that ---in the weak pumping regime--- the linear loss coefficient 
in the KdV equation results in a decay rate of the dark soliton which is twice as large 
compared to the one found in Ref.~\cite{ofy} for large-amplitude dark solitons.

The KdV model is used to describe the analytical form and the dynamics of 
dark soliton solutions supported in the polariton condensate. It is also found 
that the evolution of the dark soliton is accompanied by the emergence of a shelf: 
this is a linear wave, in the form of a long propagating tail adjacent to the soliton, 
which arises naturally in the case of dissipative KdV solitons~\cite{newell,karmas,kanew,ko,kodab}, 
as well as dissipative nonlinear Schr{\"o}dinger (NLS) 
dark solitons~\cite{dsmja,noel}. Our analytical predictions are in fairly good agreement with 
direct numerical simulations.

The paper is organized as follows. In Sec.~\ref{sec:model} we present the model and 
apply the reductive perturbation method to derive the effective KdV equation. 
Next, in Sec.~\ref{sec:dynamics}, we discuss the form and evolution of dark solitons 
and present results of direct numerical simulations. Finally, in Sec.~\ref{sec:conclu}, 
we summarize our findings and discuss interesting directions for future studies.

\section{The model and its analytical consideration}
\label{sec:model}

\subsection{The open-dissipative Gross-Pitaevskii model}

We consider an incoherently, far off-resonantly pumped exciton-polariton condensate 
in an 1D setting. In the framework of mean-field theory, 
this system can be described by means of a generalized open-dissipative GP 
equation for the macroscopic wavefunction $\Psi(x,t)$ of the polariton condensate 
coupled to a rate equation for the exciton reservoir density $n_R(x,t)$~\cite{wc,wc2,wc3}:
\begin{eqnarray}
i\hbar \frac{\partial \Psi}{\partial t} &=& -\frac{\hbar^2}{2M}\frac{\partial^2 \Psi}{\partial x^2}
+g_C |\Psi|^2\Psi + g_R n_R \Psi + \frac{i\hbar}{2}(Rn_R-\gamma_C)\Psi, 
\label{Psi} \\
\frac{\partial n_R}{\partial t} &=& P(x,t)-(\gamma_R+R|\Psi|^2)n_R. 
\label{nR}
\end{eqnarray}
Here, $M$ is the effective mass of lower polaritons, $g_C$ is the polaritons' 
interaction strength, $g_R$ is the condensate coupling to the reservoir, $R$ is the 
rate of stimulated scattering from the reservoir to the condensate, $\gamma_C$ and $\gamma_R$ 
are the polariton and exciton loss rates, respectively, while $P(x,t)$ is the exciton 
creation rate determined by the laser pumping profile. Note that in such a 1D setting, 
the transverse profiles of the densities $|\Psi|^2$ and $n_R$ are assumed to be Gaussian, 
of width $d$, determined by the thickness of the nanowire; as a result, parameters 
$g_C$, $g_R$ and $R$ assume a quasi-1D form, i.e., 
$(g_{C,R},~R) \rightarrow (g_{C,R},~R)/\sqrt{2\pi d}$~\cite{dsthv1,bob}, 
similarly to the situation occurring in the context of quasi-1D atomic BECs~\cite{siambook}.
It should also be pointed out that, in the above model, the polariton condensate is 
characterized by a repulsive (defocusing) nonlinearity, inherited from 
the repulsively interacting excitons~\cite{rmp1st,cc}.

The system of Eqs.~(\ref{Psi})-(\ref{nR}) can be expressed in a dimensionless 
form as follows: measuring $x$ in units of the healing length $\xi=\hbar/\sqrt{Mg_C n_C}$ 
(where $n_C$ is a characteristic value of the condensate density), $t$ in units of 
$t_0=\xi/c_S =\hbar/g_C n_C$ (where $c_S=\sqrt{g_C n_C/M}$ is the speed of sound), 
and densities $|\Psi|^2$ and $n_R$ in units of $n_C$.
Then, Eqs.~(\ref{Psi})-(\ref{nR}) take the form:
\begin{eqnarray}
i\frac{\partial \Psi}{\partial t} &=& -\frac{1}{2}\frac{\partial^2 \Psi}{\partial x^2}
+|\Psi|^2\Psi + g_R n_R \Psi + \frac{i}{2}(Rn_R-\gamma_C)\Psi, 
\label{Psin} \\
\frac{\partial n_R}{\partial t} &=& P(x,t)-(\gamma_R+R|\Psi|^2)n_R, 
\label{nRn}
\end{eqnarray}
where $g_R$ and $R$ are measured in units of $g_C$ and $g_C/\hbar$, respectively,   
$\gamma_C$ and $\gamma_R$ are measured in units of $1/t_0$, while the laser pump $P(x,t)$ 
is measured in units of $n_C/t_0$.

We now use the Madelung transformation $\Psi = \sqrt{\rho}\exp(i\varphi)$, 
with $\rho,~\varphi \in \mathbb{R}$, and separate real 
and imaginary parts to derive from Eqs.~(\ref{Psin})-(\ref{nRn}) the following 
real-valued system:
\begin{eqnarray}
&&\frac{\partial \rho}{\partial t}
+\frac{\partial}{\partial x}\left(\rho\frac{\partial \varphi}{\partial x}\right)
-(R n_R -\gamma_C)\rho=0, 
\label{rhot} \\ 
&&\frac{\partial \varphi}{\partial t}+\rho+\frac{1}{2}\left(\frac{\partial \varphi}{\partial x}\right)^2
-\frac{1}{2}\rho^{-1/2}\left(\frac{\partial^2 \rho^{1/2}}{\partial x^2}\right)+g_R n_R=0, 
\label{phit} \\
&&\frac{\partial n_R}{\partial t}-P+(\gamma_R +R \rho)n_R =0.
\label{nRt}
\end{eqnarray}
Next, we consider the case of a continuous-wave (cw) and spatially and temporally uniform 
pumping, i.e., $P(x,t)=P_0$, and seek for a homogeneous steady-state solution of the 
system~(\ref{rhot})-(\ref{nRt}) of the form:
\begin{eqnarray}
\rho=\rho_0, \quad n_R = n_0, \quad \varphi=-\mu t, 
\label{ss}
\end{eqnarray}  
where the unknown condensate and reservoir densities, $\rho_0$ and $n_0$, as well as 
the chemical potential $\mu$, are determined as follows. First, Eq.~(\ref{rhot}) provides 
the reservoir density:
\begin{equation}
n_0=\frac{\gamma_C}{R}.
\label{n0}
\end{equation}
Using the above result, Eq.~(\ref{phit}) leads to an equation connecting the chemical 
potential with the condensate density:
\begin{equation}
\mu = \rho_0 + \frac{g_R \gamma_C}{R}. 
\label{mu}
\end{equation}
Finally, employing Eq.~(\ref{n0}), Eq.~(\ref{nRt}) provides the condensate density:
\begin{equation}
\rho_0=\frac{1}{\gamma_C}\left(P_0-P_0^{({\rm th})}\right), 
\quad {\rm where} \quad
P_0^{({\rm th})}\equiv \frac{\gamma_R \gamma_C}{R}.
\label{rho0}
\end{equation}
Obviously, the inequality $P_0>P_0^{({\rm th})}$ must hold, so that the density 
$\rho_0$ is positive. In other words, the condensate emerges only if the uniform pump 
$P_0$ exceeds the threshold value $P_0^{({\rm th})}$. This result, as well as the 
form of the steady state solution of Eqs.~(\ref{n0})--(\ref{rho0}), are in accordance
with the analysis of Ref.~\cite{wc} (see also Ref.~\cite{ofy}). 
Here, it is also relevant to note that, having introduced $P_0^{({\rm th})}$, it is 
convenient to rewrite the equilibrium condensate density as $\rho_0=(\gamma_R/R)\alpha$, 
where the parameter $\alpha$ expresses the relative deviation of the uniform pumping $P_0$ 
from the threshold value $P_0^{({\rm th})}$, namely:
\begin{equation}
\alpha=\frac{P_0}{P_0^{({\rm th})}}-1 > 0.
\label{alpha}
\end{equation}

We now consider the physically relevant situation where the reservoir is able 
to adiabatically follow the condensate dynamics, i.e., $\gamma_C \ll \gamma_R$ \cite{wc}. 
To quantitatively define the relative magnitude of the polariton and exciton loss rates, 
we introduce a formal small parameter, $0<\epsilon \ll 1$ and assume that 
$\gamma_C=\epsilon \tilde{\gamma}_C$, where $\tilde{\gamma}_C$, as well as 
$\gamma_R$, are taken to be of order $\mathcal{O}(1)$. 
Furthermore, it is assumed that the scattering rate, $R$, of reservoir particles into the
condensate, as well as the relative deviation of the pumping from the threshold, $\alpha$, 
are sufficiently small \cite{ofy}, i.e., $R = \epsilon \tilde{R}$ and 
$\alpha = \epsilon \tilde{\alpha}$ [where $\tilde{R}$ and $\tilde{\alpha}$ are 
of order $\mathcal{O}(1)$]. Thus, the relative magnitude (and smallness) 
of all physical parameters involved in the problem, defined through 
the formal small parameter $\epsilon$, is summarized as follows:
\begin{equation}
\alpha = \epsilon \tilde{\alpha}, \quad
\gamma_C=\epsilon \tilde{\gamma}_C, \quad 
R=\epsilon \tilde{R}, \quad 
\gamma_R=\mathcal{O}(1), \quad 
g_R=\mathcal{O}(1). 
\label{wp}
\end{equation}
It is worth mentioning that, under the above assumptions, all steady state 
parameters, namely the densities $\rho_0$ and $n_0$, the pump threshold $P_0^{({\rm th})}$, 
as well as the chemical potential $\mu$, are of order $\mathcal{O}(1)$.


\subsection{Reductive perturbation method and Korteweg-de Vries equation}

Next, considering a perturbative regime of sufficiently small reservoir excitations, 
we will use the reductive perturbation method \cite{rpm} to determine 
small-amplitude and slowly-varying modulations of the steady state. 
We thus seek solutions of Eqs.~(\ref{rhot})--(\ref{nRt}) in the form of the 
asymptotic expansions:
\begin{eqnarray}
\rho&=&\phantom{-}\rho_0 + \epsilon \rho_1(X,T)+ \epsilon^2 \rho_2(X,T)+ \cdots,  
\label{pert1} \\
\varphi&=&-\mu t + \epsilon^{1/2} \varphi_1(X,T)+\epsilon^{3/2} \varphi_2(X,T)+\cdots,   
\label{pert2} \\
n_R&=& \phantom{-}n_0 + \epsilon^2 n_1(X,T)+\epsilon^3 n_2(X,T)+\cdots,    
\label{pert3}
\end{eqnarray}
where the unknown real functions $\rho_j$, $\varphi_j$ and $n_j$ ($j=1,2,\ldots$) depend 
on the slow variables:
\begin{equation}
X=\epsilon^{1/2} (x-vt), \quad T=\epsilon^{3/2}t, 
\end{equation}
with $v$ being an unknown velocity, to be determined self-consistently at a 
later stage of our analysis (see below). 
We now substitute the expansions~(\ref{pert1})--(\ref{pert3}) 
into Eqs.~(\ref{rhot})--(\ref{nRt}), 
and taking into account the scaling of the parameters [cf. Eq.~(\ref{wp})], 
we equate terms of the same order 
in $\epsilon$, and obtain the following results. 

At leading order in $\epsilon$, namely at orders 
$\mathcal{O}(\epsilon^{3/2})$ and $\mathcal{O}(\epsilon)$, Eqs.~(\ref{rhot})-(\ref{phit}) 
lead, respectively, to the following linear system: 
\begin{eqnarray}
-v\frac{\partial \rho_1}{\partial X}+\rho_0 \frac{\partial^2 \varphi_1}{\partial X^2}&=&0, 
\label{linearA}
\\[1.0ex]
-v\frac{\partial \varphi_1}{\partial X}+\rho_1&=&0.
\label{linearB}
\end{eqnarray}
Its compatibility condition is the algebraic equation:
\begin{equation}
v^2=\rho_0,
\label{v}
\end{equation}
which determines the velocity $v$ of the linear 
excitations propagating on top of the cw background, also referred to as the speed of sound $v$. 
In addition, Eq.~(\ref{nRt}), at the leading order in $\epsilon$, i.e., at order 
$\mathcal{O}(\epsilon^{2})$, leads to the equation:
\begin{equation}
n_1 = -\frac{\tilde{\gamma}_C}{\gamma_R} \rho_1, 
\label{nqconn}
\end{equation}
connecting the reservoir density $n_1$ to the polariton density $\rho_1$. Obviously, 
once $\rho_1$ is found, then $\varphi_1$ and $n_1$ can be respectively derived from 
Eqs.~(\ref{linearA}), (\ref{linearB}), and (\ref{nqconn}). 
We thus proceed to the next order in $\epsilon$, 
and derive from Eqs.~(\ref{rhot}) and (\ref{phit}), at orders 
$\mathcal{O}(\epsilon^{5/2})$ and $\mathcal{O}(\epsilon^2)$ respectively,  
the following nonlinear equations: 
\begin{eqnarray}
&&\frac{\partial \rho_1}{\partial T}-v\frac{\partial \rho_2}{\partial X}
+\rho_0 \frac{\partial^2 \varphi_2}{\partial X^2}
+\frac{\partial}{\partial X}\left(\rho_1\frac{\partial \varphi_1}{\partial X}\right)
=0,
\label{nl1} \\
&&\frac{\partial \varphi_1}{\partial T}-v\frac{\partial \varphi_2}{\partial X}+\rho_2
+\frac{1}{2}\left(\frac{\partial \varphi_1}{\partial X}\right)^2
-\frac{1}{4\rho_0}\frac{\partial^2 \rho_1}{\partial X^2}+g_R n_1=0.
\label{nl2} 
\end{eqnarray}
The compatibility condition for the above equations can be found as follows. 
First, use Eq.~(\ref{linearB}), as well as Eq.~(\ref{nqconn}) to 
express $\partial \varphi_1/\partial x$ and $n_1$ in terms of $q_1$. Next, 
differentiate Eq.~(\ref{nl2}) once with respect to $X$, multiply by $v$, and add to 
the resulting equation Eq.~(\ref{nl1}). The compatibility condition for 
Eq.~(\ref{nl1})-(\ref{nl2}) is the above mentioned algebraic equation~(\ref{v}), 
together with the following KdV equation:
\begin{equation}
\frac{\partial \rho_1}{\partial T}
-\frac{v g_R \tilde{\gamma}_C}{2\gamma_R} \frac{\partial \rho_1}{\partial X} 
+ \frac{3}{2v} \rho_1 \frac{\partial \rho_1}{\partial X} 
-\frac{v}{8\rho_0} \frac{\partial^3 \rho_1}{\partial X^3}=0, 
\label{kdv1}
\end{equation}
At the present order of approximation, the above model does not incorporate dissipative 
terms. The lowest order for such term appears in Eq.~(\ref{rhot}), and has the form 
$-\epsilon^3 \tilde{\alpha}\gamma_R n_1$, i.e., it is a term of order $\mathcal{O}(\epsilon^3)$. 
Then, to take into account this term, we may modify Eq.~(\ref{nl1}) by adding 
to its right-hand side the additional term $\epsilon^{1/2} \tilde{\alpha}\gamma_R n_1$. 
This, in turn, amounts to incorporating the term 
$-(\epsilon^{1/2}/2)\tilde{\alpha} \tilde{\gamma}_C \rho_1$
to the right-hand side of the KdV Eq.~(\ref{kdv1}). 
To this end, taking into regard this modification, 
we proceed by expressing the KdV equation in its standard dimensionless form \cite{ablowitz2}. 
First, we apply to Eq.~(\ref{kdv1}) the Galilean transformation 
$X=\chi+[(vg_R \tilde{\gamma}_C)/(2\gamma_R)]T$ 
to remove 
the first-order spatial derivative term. Next, employing the scale transformations 
$\tau=-(v/8\rho_0)T$
and $\rho_1=(1/2)u$, we obtain:
\begin{equation}
\frac{\partial u}{\partial \tau}-6u\frac{\partial u}{\partial \chi}
+\frac{\partial^3 u}{\partial \chi^3 }=\Gamma u,
\label{kdv}
\end{equation}
where $\Gamma= 4 \epsilon^{1/2}\tilde{\alpha}\tilde{\gamma}_C \rho_0/v$. 
Notice that the above equation is, in fact, a KdV equation with linear loss. 
This model plays a key role 
in our analysis; it is used below to determine both the analytical form 
and the dynamics of approximate (small-amplitude) dark soliton solutions 
that can be supported in polariton superfluids. 

\section{Dynamics of dark solitons}
\label{sec:dynamics}

\subsection{Analytical results}

Based on the connection between the open dissipative GP 
model (\ref{Psin})-(\ref{nRn}) and the KdV Eq.~(\ref{kdv}), we can 
use solutions of the latter to construct approximate solutions of the original problem. 
Thus, if $u$ satisfies Eq.~(\ref{kdv}) then: 
\begin{eqnarray}
\Psi &\approx& \sqrt{\rho_0+\frac{1}{2}\epsilon u} 
\,\exp\left(-i\mu t+i\frac{\epsilon^{1/2}}{2v} \int_{-\infty}^{\chi} u{\rm d}\chi' \right),
\label{polsol} \\
n_R &\approx& n_0 -\epsilon \frac{\gamma_C}{2\gamma_R}u.
\label{nf}
\end{eqnarray}

Let us now examine separately the cases with $\Gamma=0$ and $\Gamma \ne 0$. 
In the former (lossless) case, $\Gamma=0$, Eq.~(\ref{kdv}) becomes the completely 
integrable KdV equation, which possesses 
the single-soliton solution, $u=u_s$, given by \cite{ablowitz2}:
\begin{eqnarray}
u_s(Z)=-2\kappa^2{\rm sech}^2(Z), \quad Z=\kappa[\chi-\eta(\tau)], 
\label{sol1}
\end{eqnarray}
where $\kappa$ is a free parameter linking the soliton's amplitude to its velocity, 
$\eta(\tau)=4\kappa^2 \tau+\eta_0$ is the soliton center 
(with the constant $\eta_0$ denoting the initial soliton location), and 
$d\eta/d\tau=4\kappa^2$ is the soliton velocity in the $(\chi,\tau)$ reference frame. 
Thus, in this case, up to order $\mathcal{O}(\epsilon)$, 
the macroscopic wavefunction $\Psi$ of the polariton condensate 
and the exciton density $n_R$, can be expressed in terms of the original (dimensionless) 
coordinates $x$ and $t$ and physical parameters as follows:
\begin{eqnarray}
\Psi &\approx& \sqrt{\rho_0 -\epsilon \kappa^2{\rm sech}^2(Z)}
\,\exp\left(-i\mu t
-i\frac{\epsilon^{1/2}\kappa}{\sqrt{\rho_0}} \tanh(Z) \right),
\label{polsol2} \\
n_R &\approx& n_0 +\epsilon \kappa^2 \frac{\gamma_C}{\gamma_R}{\rm sech}^2(Z),
\label{nf2} \\
Z&=&\epsilon^{1/2}\kappa\left\{x-\sqrt{\rho_0}
\left[\left(1-\frac{\gamma_C g_R}{2\gamma_R}\right)
-\epsilon \frac{\kappa^2}{2\rho_0} \right]t -x_0 \right\},
\label{Z}
\end{eqnarray}
where $x_0$ is the initial soliton position, and we have considered, 
without loss of generality, right-going waves with $v=\sqrt{\rho_0}$.
Clearly, the solution~(\ref{polsol}) has the form of a sech-shaped 
density dip, with a tanh-shaped phase jump across the density minimum, and it 
is thus a dark soliton. On the other hand, the exciton 
density~(\ref{nf}) follows the form of an anti-dark soliton soliton, i.e., it has a
sech$^2$ hump shape on top of the background, 
at the location of the dark polariton soliton, and asymptotes (for $x\rightarrow \pm \infty$) 
to the equilibrium density $n_0$. 

Next, we turn to the $\Gamma \ne 0$ case to study the role of the linear loss 
on the soliton dynamics. 
First we note that, as is known, the KdV equation was first derived to describe 
the evolution of shallow water waves~\cite{ablowitz2}. Nevertheless, in 
the case where the water's depth is nonuniform, the KdV incorporates an effective 
dissipative perturbation of the form $\Gamma u$, as in the case of Eq.~(\ref{kdv}), 
with $\Gamma$ being proportional to the (small) gradient of the water's depth~\cite{newell}. 
Interestingly, in our case, $\Gamma$ is connected with parameters characterizing the 
open-dissipative nature of the problem (such as the polariton 
loss rate $\gamma_C$), a fact establishing an interesting connection 
of the polariton superfluids problem with the one of shallow water waves. 

The problem of the KdV soliton dynamics in the case $\Gamma\ne 0$ has been analyzed 
in the past in various works, using a perturbed inverse scattering transform (IST) 
theory~\cite{karmas,kanew} and asymptotic expansion methods~\cite{ko,kodab} (see also 
the review~\cite{kivmal}). 
The main results of the analyses reported in these works are as follows. 
For sufficiently small $\Gamma$ [as in our case, where $\Gamma=\mathcal{O}(\epsilon^{1/2})$], 
the solution of Eq.~(\ref{kdv}) can be expressed in the form:
\begin{equation}
u(\chi,\tau)=u_s(Z;\kappa) + \delta u(Z,\tau), 
\label{persol}
\end{equation}
where $u_s$ and $\delta u$ denote, respectively, the soliton and the radiative 
components of the solution. The soliton component has the functional form given in 
Eq.~(\ref{sol1}), but the parameters setting the amplitude and velocity of the soliton, 
$\kappa$ and $d\eta/d\tau$, become functions of time. In particular, their evolution 
is given by the expressions:
\begin{equation}
\kappa(\tau)=\kappa(0)\exp\left(\frac{2}{3}\Gamma \tau\right), 
\quad
\frac{d\eta}{d\tau}=4 \kappa^2(\tau)+\frac{\Gamma}{3\kappa(\tau)},
\label{kapa}
\end{equation}
with the second of the above equations leading to the result:
%
\begin{eqnarray}
    \eta(\tau)=\frac{1}{2\kappa(0)} \left\{ \frac{6\kappa^3(0)}{\Gamma} 
    \left[ \exp\left(\frac{4}{3}\Gamma\tau\right)-1 \right] - \left[\exp\left(-\frac{2}{3}\Gamma\tau\right) -1\right]  \right\}
    +\eta_0 ,
    \label{deta}
\end{eqnarray}
where once again $\eta_0$ denotes the initial soliton position.
The evolution equation~(\ref{kapa}) 
indicates that the dark soliton's amplitude 
decays exponentially in time: indeed, the exponential law $\exp(2\Gamma \tau/3)$, 
when transformed back in the original time takes the form $\exp(-t/t_\star)$, 
where the soliton decay rate $t_\star$ is given by: 
\begin{equation}
t_\star = \frac{3}{\alpha \gamma_C} t_0
=\frac{3}{\gamma_C} \frac{P_0^{({\rm th})}}{P_0-P_0^{({\rm th})}} ~t_0,  
\label{lifetime}
\end{equation} 
where $t_0$ is the characteristic time scale for the system introduced in Sec.~\ref{sec:model}. 
It is observed that the soliton's decay rate in the weak pumping regime 
depends on the decay rate $\gamma_C$ of the polariton condensate, as well as 
the relative deviation $\alpha$ of the uniform pumping $P_0$ 
from the threshold value $P_0^{({\rm th})}$. This decay rate is twice 
as large compared to the one found in Ref.~\cite{ofy}. This means that 
the small-amplitude dark solitons hereby predcited predicted decay faster than the 
large-amplitude ones considered in Ref.~\cite{ofy}.

In the present case of $\Gamma \ne 0$, there exists also 
the radiation component $\delta u$, which is emitted by the soliton under 
the action of the perturbation. This component has the form of a {\it shelf}, 
whose generation was studied by means of the IST perturbation theory and asymptotic 
expansion techniques \cite{karmas,kanew,ko,kodab}. 
According to these works, the structure of the shelf can be found 
in a closed analytical form which, however, is cumbersome to be presented here. 
Instead, we present the asymptotic form of the shelf:
\begin{eqnarray}
\delta u \approx
\begin{array}{ll}
 -\frac{2\Gamma}{3\kappa(\tau)}Z^2\exp(-2Z), & Z\gg 1, 
\nonumber 
\\
\nonumber 
-\frac{\Gamma}{3\kappa(\tau)}\left[1+2Z^2\exp(2Z) \right], & -Z \gg 1.
\end{array}
\end{eqnarray}
Notice that at the times $\kappa^{-2}(0) \ll \tau \ll \Gamma^{-1}$ 
(recall that $\kappa(0)$ is the initial soliton amplitude) the form of the shelf 
is even simpler \cite{kivmal}: 
in the region $0 < \chi < \mathcal{D}(\tau) \equiv \int_{0}^{\tau}4\kappa^2(\tau'){\rm d}\tau'$, 
the wave field is approximately uniform, namely:
\begin{equation}
\delta u \approx -\frac{\Gamma}{3\kappa(\tau)},
\label{shelf}
\end{equation}
while outside this region $u$ may be set equal to zero. 
Here, the initial coordinate of the soliton is $\chi_0=0$, 
while $\mathcal{D}(\tau)$ is the distance traveled by the soliton up
to the moment $\tau$, and $\kappa(\tau)$ evolves in time according to
Eq.~(\ref{kapa}). The above asymptotic result leads ---according to Eq.~(\ref{polsol})--- 
to a simple estimation of the shelf amplitude in terms of the original variables, namely: 
\begin{equation}
|\Psi|^2-\rho_0 \approx \frac{2}{3} \epsilon^{3/2}
\frac{\tilde{\alpha}\tilde{\gamma}_C}{\kappa(\tau)\sqrt{\rho_0}}.
\label{sh2}
\end{equation}
Thus, under the action of the perturbation, the original soliton 
changes speed and shape, and forms a shelf ---namely a long, almost constant  
(sufficiently far away from the soliton) tail accompanying the soliton, 
at the end of which there are small oscillations in time and space.
Notice that the emergence of the shelf in the problem under consideration, is in accordance 
with the analysis of Ref.~\cite{dsmja}: in this work, a multiscale boundary layer perturbation 
theory was used to show that shelves appear generically when dark solitons evolve under the 
action of dissipative perturbations (see also Ref.~\cite{noel} for an analysis on 
dark solitons and shelves in nonlocal media).

\subsection{Numerical results}

\begin{figure}[tbp]
	\centering
	\includegraphics[width=0.7\linewidth]{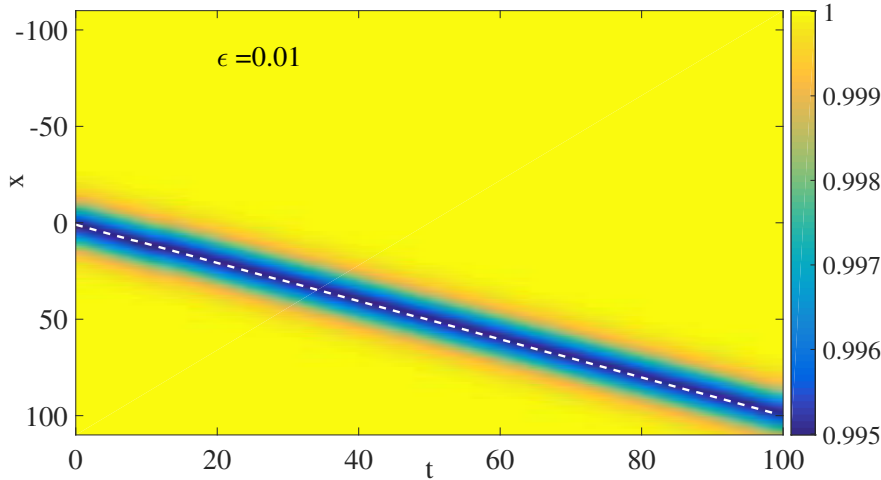} 
	\caption{
		Propagation of the polariton wavefunction modulus, for $\epsilon=0.01$. The dashed (white) 
		line shows the theoretical prediction of our model for the position of the soliton center 
		(where the density is minimum), for the same initial profile.
	}
	\label{fig:FIG2_eps_01} 
\end{figure}

Next, we test the validity of our approximations and analytical results by means of direct 
numerical simulations for the system of Eqs.~(\ref{Psin})-(\ref{nRn}). Our aim is to compare 
these results with our analytic predictions presented above. 
For simplicity, in our simulations, all quantities with tildes have been set equal to one, 
so that our initial data depend on a sole parameter, namely the small parameter $\epsilon$. 
Since our approach relies on a perturbation expansion, 
we expect results for relatively smaller values of $\epsilon$ to give a better agreement. 

\begin{figure}[tbp]
	\centering
	\includegraphics[width=0.7\linewidth]{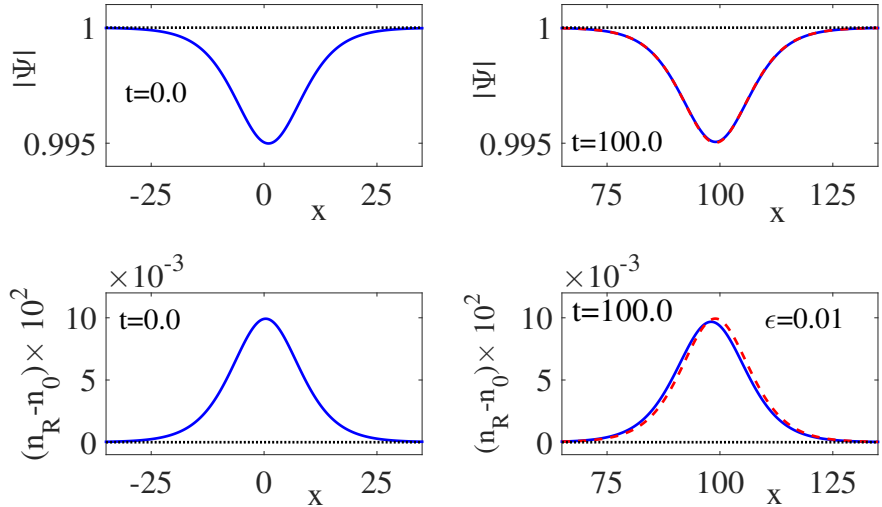}
	\caption{
		Initial (left panels) and final (right panels) ($t=100$) polariton wavefunction 
		modulus (top panels) and reservoir exciton density (bottom), for $\epsilon=0.01$. 
		The solid (blue) and dashed (red) lines 
		show, respectively, the numerical and the theoretical predictions of our model, 
		for the same initial profile, namely the KdV soliton.
	}
	\label{fig:FIG1_eps_01} 
\end{figure}

\begin{figure}[htbp]
	\centering
	\includegraphics[width=0.7\linewidth]{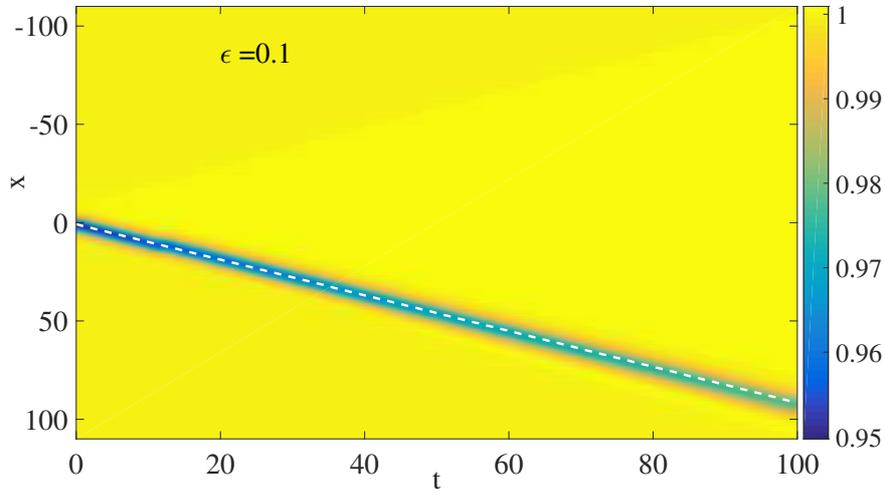} 
	\caption {
		Same as in Fig.~\ref{fig:FIG2_eps_01} for $\epsilon=0.1$.
	}
	\label{fig:FIG4_eps_1} 
\end{figure}

\begin{figure}[htbp]
	\centering
	\includegraphics[width=0.7\linewidth]{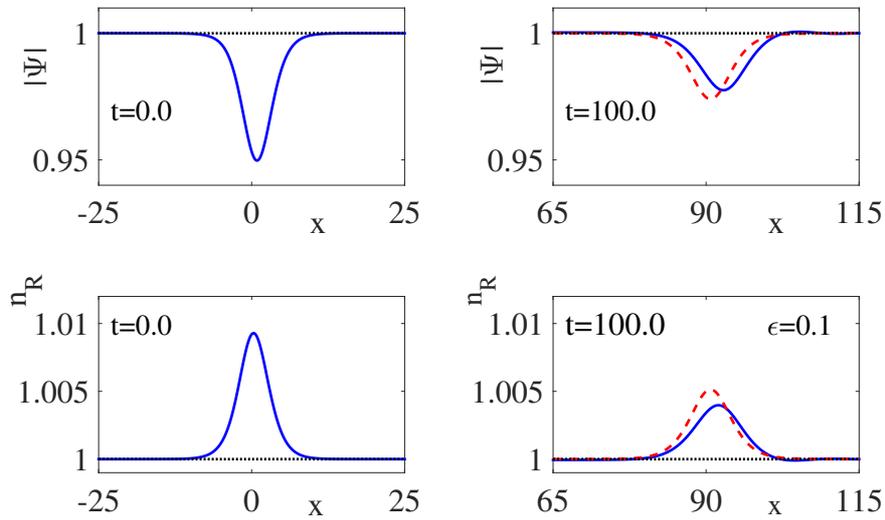} 
	\caption {
		Same as in Fig.~\ref{fig:FIG1_eps_01} for $\epsilon=0.1$.
	}
	\label{fig:FIG3_eps_1} 
\end{figure}

A series of illustrative results are shown in Figs.~\ref{fig:FIG2_eps_01}--\ref{fig:FIG3_eps_1}. 
In particular, Fig.~\ref{fig:FIG2_eps_01} depicts, for $\epsilon=0.01$, the full evolution of
the polariton's wavefunction modulus. Here, it is observed that the analytical prediction 
for the soliton trajectory, corresponding to the dashed (red) line, 
closely follows the numerical result. In addition, to further verify our estimates for the 
soliton's decay rate and velocity, we depict in Fig.~\ref{fig:FIG1_eps_01} the 
initial (left) and final (right) profiles for $|\Psi|$ (top panels) and $n_R$ (bottom panels), 
in the same case of $\epsilon=0.01$. We also depict in the panels, shown with a dashed 
(red) line, the prediction of our reductive perturbation theory. 
For this value of the small parameter, the two are nearly indistinguishable.

Next, we use a relatively large value of the small parameter, namely $\epsilon=0.1$. 
As was to be expected, the deviations for the same duration of the propagation are more evident. 
First, the contour plot for this scenario, depicted in Fig.~\ref{fig:FIG4_eps_1}, 
shows again a fairly good agreement between our model and the numerics of the full equations,  
at least up to $t\approx 50$. Nevertheless, as it is shown in Fig.~\ref{fig:FIG3_eps_1}, 
where the initial and final (at $t=100$) snapshots for $|\Psi|$ and $n_R-n_0$ are depicted, 
the analytic results underestimate the speed (but only by a few percent) and overestimate 
the strength of the peaks (by some $15\%$). 

\begin{figure}[tbp]
	\centering
	\includegraphics[width=0.7\linewidth]{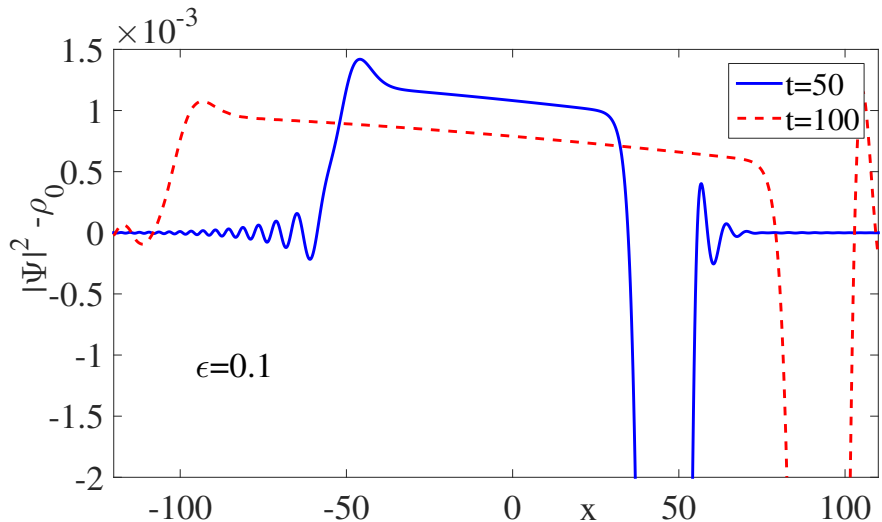} 
	\caption{Linear radiation (shelf) of the polariton wavefunction modulus 
	at $t=50$ (solid blue line) and $t=50$ (dashed red line), for $\epsilon=0.1$. 
	The `dips' of the main dark soliton pulses can be clearly observed.
	}
	\label{fig:FIG5_shelf} 
\end{figure}

For the the same case of $\epsilon=0.1$, Fig.~\ref{fig:FIG5_shelf} depicts 
the emergence and evolution of the shelf. It is clear that, in accordance to our 
analytical predictions based on the KdV theory, 
the shelf is on top of the background density, and its front propagates in the 
direction opposite to that of the soliton, at roughly the same speed. 
It is observed that the shape of the shelf becomes flatter as it propagates, 
and at the end of the shelf, small oscillations are clearly observed, 
which is the typical scenario in the KdV dynamics~\cite{karmas,kanew,ko,kodab}. 
Thus, the results of our simulations are in a qualitative agreement with the 
effective KdV picture. Nevertheless, we should note that the prediction of Eq.~(\ref{sh2}) 
fails to quantitatively capture the size of the shelf (the analytical estimate is  
about an order of magnitude larger than the numerical result), a fact that 
can be attributed to the asymptotic nature of the analytical prediction.

\begin{figure}[htbp]
 	\centering
 	\includegraphics[width=0.7\linewidth]{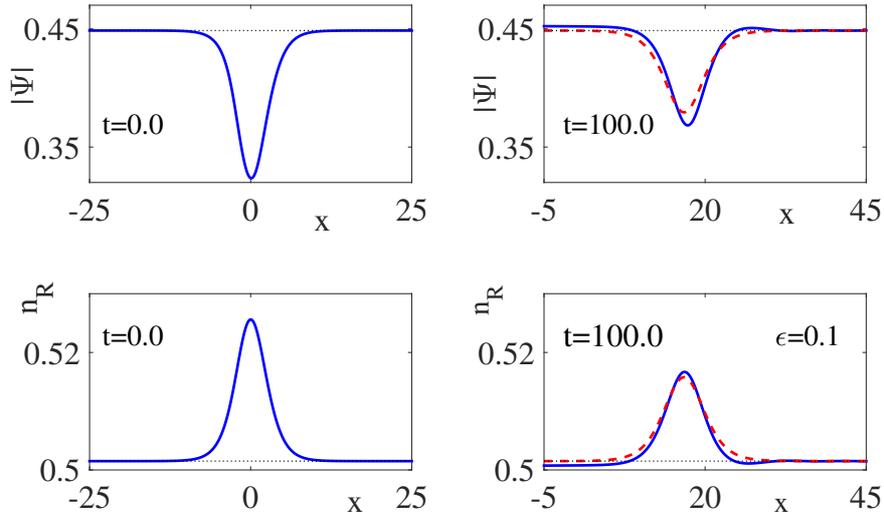} 
 	\caption{Same as Fig.\ref{fig:FIG3_eps_1}, but for parameters 
 		$\tilde{g}_R=0.33$,
 		$\tilde{\gamma}_C=1.67$, $\tilde{\gamma}_R=0.67$, and $\tilde{R}=3.33$. 
 	}
 	\label{fig:FIG6_phys} 
 \end{figure}

As a final illustration we report the results for a scenario corresponding to physically attainable
parameter values -- cf., e.g.,  
data presented in Ref.~\cite{wc2}. First we note that both parameters 
$\tilde{g}_R$ and $R$ are not very well known: 
the first is frequently set to zero, while the second
can be made to vary (for instance, by playing with the size of the
confinement, or width of the quantum well). 
Here we select values of $g_R=0.005\, \mathrm{meV \cdot \mu m^2}$ and $\hbar R = 0.005\,
 \mathrm{meV \cdot \mu m^2}$, together with $g_c=3 g_R=0.015\, \mathrm{meV \cdot \mu m^2}$,
 $\hbar \gamma_R = 4\times \hbar \gamma_C = 2\, \mathrm{meV}$. 
 From these, and the choice $n_C=2\times 10^{14}\, \mathrm{m}^{-2}$, and for the case of $\epsilon=0.1$, the scaled (tilde) values are $\tilde{\alpha}=1$, $\tilde{g}_R=0.33$,
 $\tilde{\gamma}_C=1.67$, $\tilde{\gamma}_R=0.67$, and $\tilde{R}=3.33$.
  For these parameters, the result of the evolution of the dark soliton is shown
 in Fig.~\ref{fig:FIG6_phys}. As we can see, the agreement between the numerical evolution and the KdV
 approximation is again fairly good, even for this more stringent value of $\epsilon$. Notice that only the initial and final profiles are shown, but this
 level of agreement is maintained throughout the course of the
 numerical simulation.
 
\section{Conclusions}
\label{sec:conclu}

We have studied an open dissipative mean-field model for 
exciton-polariton condensates. In particular, the considered system is composed 
of a generalized open-dissipative Gross-Pitaevskii equation, describing the macroscopic 
wavefunction of an incoherently pumped polariton condensate, coupled to a rate equation 
for the exciton reservoir density. Considering a uniform pumping, and assuming that 
the polariton loss rate and the rate of the stimulated scattering into the condensate 
are sufficiently small, we have analyzed the regime of weak pumping.

Using a perturbative approach, we have derived an effective KdV equation with linear 
loss. This KdV model was used 
to describe the analytical form and the dynamics of approximate 
dark soliton solutions that can be supported in exciton-polariton condensates. Thus, 
it was found that the polariton field supports a decaying dark soliton, with a decay rate 
depending on the physical parameters of the problem, such as the polariton decay rate and 
the relative deviation of the uniform pumping from its threshold value. 
It was also found that the evolution 
of the dark soliton is accompanied by a shelf, whose emergence and subsequent dynamics are 
in qualitative agreement with the KdV picture. The analytical findings 
were found to be in fairly good agreement with the 
direct numerical simulations, even for the case of a relatively 
large value of the formal small parameter. 

It would be interesting to extend our considerations to multi-dimensional and multi-component 
(spinor) polariton superfluid settings ---see, e.g. Refs.~\cite{spinor,berl3} and 
Refs.~\cite{dsth3,dsthv2} for work on spin dynamics of dark polariton solitons. 
In this setting, a quite relevant investigation 
would concern the existence of spinorial, vortex-free dark solitonic structures in such systems;  
notice that such states are known to exist in single-component 
Gross-Pitaevski/nonlinear Schr{\"o}dinger systems, 
where small-amplitude dark solitonic structures obey effective Kadomtsev-Petviashvilli (KP) 
equations ---see, e.g., the review \cite{kivpeli} and references therein. 
It should also be interesting to use the methodology devised in this work to study 
other models that are used in the context of open dissipative systems, such as the 
Lugiato-Lefever 
equation~\cite{ll} describing dissipative dynamics in optical resonators, and also 
supports ---in the defocusing regime--- dark solitons~\cite{kno}.

\hspace{+5pt}

{\bf Acknowledgements.}
P.G.K. gratefully acknowledges the support of NSF-PHY-1602994 and of the Stavros
Niarchos Foundation via the Greek Diaspora Fellowship Program.
P.G.K. and A.S.R. also acknowledge the hospitality of the Synthetic Quantum Systems group and
of Markus Oberthaler at the Kirchhoff Institute for Physics
(KIP) at the University of Heidelberg, as well as that of the
Center for Optical Quantum Technologies (ZOQ) and of Peter
Schmelcher at the University of Hamburg
A.S.R. acknowledges financial support from FCT throught grant UID/FIS/04650/2013.
R.C.G.~acknowledges support from NSF-PHY-1603058.


\end{document}